\documentstyle[11pt]{article}

\begin{document}
\setlength{\parindent}{5 mm}

\catcode`\@=11
\renewcommand{\theequation}{\thesection.\arabic{equation}}
\@addtoreset{equation}{section}
\catcode`\@=10

\def\dsp{\displaystyle}
\def\d{\mbox{\rm d}}
\def\e{\mbox{\rm e}}

\def\r{\prime}
\def\p{\partial}
\def\ut#1{{\mathunderaccent\tilde #1}}
\def\cosec{{\rm cosec}}
\def\kd{{\delta _{ij}}}
\def\ke{{\epsilon _{ijk}}}
\def\ha{\mbox{$\frac{1}{2}$}}
\def\tha{\mbox{$\frac{3}{2}$}}
\def\fha{\mbox{$\frac{5}{2}$}}
\def\nha{\mbox{$\frac{n}{2}$}}
\def\oq{\mbox{$\frac{1}{4}$}}
\def\tq{\mbox{$\frac{3}{4}$}}
\def\tth{\mbox{$\frac{2}{3}$}}
\def\mod{\mbox{\rm mod}}
\def\f{\frac}
\def\p{\dsp\partial}
\def\arcsinh{{\rm arcsinh}}
\def\sinh{{\rm sinh}}
\def\cosh{{\rm cosh}}
\def\arccosh{{\rm arccosh}}
\def\arctanh{{\rm arctanh}}
\def\tanh{{\rm tanh}}
\def\coth{{\rm coth}}
\def\sech{{\rm sech}}
\def\cosech{{\rm cosech}}


\def\ga{\alpha}
\def\gb{\beta}
\def\gc{\chi}
\def\gd{\delta}
\def\ge{\epsilon}
\def\gf{\phi}
\def\gg{\gamma}
\def\gi{\iota}
\def\gk{\kappa}
\def\gl{\lambda}
\def\gp{\psi}
\def\gs{\sigma}
\def\gt{\theta}
\def\gu{\upsilon}
\def\gve{\varepsilon}
\def\gw{\omega}
\def\gz{\zeta}
\def\gG{\Gamma}
\def\gD{\Delta}
\def\gF{\Phi}
\def\gL{\Lambda}
\def\gP{\Psi}
\def\gS{\Sigma}
\def\gT{\Theta}
\def\gW{\Omega}

\def\bhw{\hat{\mbox{\boldmath $\omega$}}}
\def\bht{\hat{\mbox{\boldmath $\theta$}}}
\def\bhf{\hat{\mbox{\boldmath $\phi$}}}

\def\bhr{\hat{\mbox{\boldmath $r$}}}
\def\bhL{\hat{\mbox{\boldmath $L$}}}


\def\bfA{{\bf A}}
\def\bfB{{\bf B}}
\def\bfC{{\bf C}}
\def\bfD{{\bf D}}
\def\bfE{{\bf E}}
\def\bfF{{\bf F}}
\def\bfG{{\bf G}}
\def\bfH{{\bf H}}
\def\bfI{{\bf I}}
\def\bfJ{{\bf J}}
\def\bfK{{\bf K}}
\def\bfL{{\bf L}}
\def\bfM{{\bf M}}
\def\bfN{{\bf N}}
\def\bfO{{\bf O}}
\def\bfP{{\bf P}}
\def\bfQ{{\bf Q}}
\def\bfR{{\bf R}}
\def\bfS{{\bf S}}
\def\bfT{{\bf T}}
\def\bfU{{\bf U}}
\def\bfV{{\bf V}}
\def\bfW{{\bf W}}
\def\bfX{{\bf X}}
\def\bfY{{\bf Y}}
\def\bfZ{{\bf Z}}
\def\bfa{{\bf a}}
\def\bfb{{\bf b}}
\def\bfc{{\bf c}}
\def\bfd{{\bf d}}
\def\bfe{{\bf e}}
\def\bff{{\bf f}}
\def\bfg{{\bf g}}
\def\bfh{{\bf h}}
\def\bfi{{\bf i}}
\def\bfj{{\bf j}}
\def\bfk{{\bf k}}
\def\bfl{{\bf l}}
\def\bfm{{\bf m}}
\def\bfn{{\bf n}}
\def\bfo{{\bf o}}
\def\bfp{{\bf p}}
\def\bfq{{\bf q}}
\def\bfr{{\bf r}}
\def\bfs{{\bf s}}
\def\bft{{\bf t}}
\def\bfu{{\bf u}}
\def\bfv{{\bf v}}
\def\bfw{{\bf w}}
\def\bfx{{\bf x}}
\def\bfy{{\bf y}}
\def\bfz{{\bf z}}


\def\p{\dsp \partial}

\def\pb{{\p\over\p b}}
\def\pc{{\p\over\p c}}
\def\pd{{\p\over\p d}}
\def\pe{{\p\over\p e}}
\def\pf{{\p\over\p f}}
\def\pg{{\p\over\p g}}
\def\ph{{\p\over\p h}}

\def\pj{{\p\over\p j}}
\def\pk{{\p\over\p k}}
\def\pl{{\p\over\p l}}

\def\pn{{\p\over\p n}}
\def\po{{\p\over\p o}}

\def\pq{{\p\over\p q}}
\def\pr{{\p\over\p r}}
\def\ps{{\p\over\p s}}
\def\pt{{\p\over\p t}}
\def\pu{{\p\over\p u}}
\def\pv{{\p\over\p v}}
\def\pw{{\p\over\p w}}
\def\px{{\p\over\p x}}
\def\py{{\p\over\p y}}
\def\pz{{\p\over\p z}}

\def\pA{{\p\over\p A}}
\def\pB{{\p\over\p B}}
\def\pC{{\p\over\p C}}
\def\pD{{\p\over\p D}}
\def\pE{{\p\over\p E}}
\def\pF{{\p\over\p F}}
\def\pG{{\p\over\p G}}
\def\pH{{\p\over\p H}}
\def\pI{{\p\over\p I}}
\def\pJ{{\p\over\p J}}
\def\pK{{\p\over\p K}}
\def\pL{{\p\over\p L}}
\def\pM{{\p\over\p M}}
\def\pN{{\p\over\p N}}
\def\pO{{\p\over\p O}}

\def\pQ{{\p\over\p Q}}
\def\pR{{\p\over\p R}}
\def\pS{{\p\over\p S}}
\def\pT{{\p\over\p T}}
\def\pU{{\p\over\p U}}
\def\pV{{\p\over\p V}}
\def\pW{{\p\over\p W}}
\def\pX{{\p\over\p X}}
\def\pY{{\p\over\p Y}}
\def\pZ{{\p\over\p Z}}

\def\da{{\d\over\d a}}
\def\db{{\d\over\d b}}
\def\dc{{\d\over\d c}}
\def\de{{\d\over\d e}}
\def\df{{\d\over\d f}}
\def\dg{{\d\over\d g}}
\def\dh{{\d\over\d h}}
\def\di{{\d\over\d i}}
\def\dj{{\d\over\d j}}
\def\dk{{\d\over\d k}}
\def\dl{{\d\over\d l}}
\def\dn{{\d\over\d n}}
\def\do{{\d\over\d o}}
\def\ddp{{\d\over\d p}}
\def\dq{{\d\over\d q}}
\def\dr{{\d\over\d r}}
\def\ds{{\d\over\d s}}
\def\\d t{{\d\over\d t}}
\def\du{{\d\over\d u}}
\def\dv{{\d\over\d v}}
\def\dw{{\d\over\d w}}
\def\dx{{\d\over\d x}}
\def\dy{{\d\over\d y}}
\def\dz{{\d\over\d z}}

\def\dA{{\d\over\d A}}
\def\dB{{\d\over\d B}}
\def\dC{{\d\over\d C}}
\def\dD{{\d\over\d D}}
\def\dE{{\d\over\d E}}
\def\dF{{\d\over\d F}}
\def\dG{{\d\over\d G}}
\def\dH{{\d\over\d H}}
\def\dI{{\d\over\d I}}
\def\dJ{{\d\over\d J}}
\def\dK{{\d\over\d K}}
\def\dL{{\d\over\d L}}
\def\dM{{\d\over\d M}}
\def\dN{{\d\over\d N}}
\def\dO{{\d\over\d O}}
\def\dP{{\d\over\d P}}
\def\dQ{{\d\over\d Q}}
\def\dR{{\d\over\d R}}
\def\dS{{\d\over\d S}}
\def\\d t{{\d\over\d T}}
\def\dU{{\d\over\d U}}
\def\dV{{\d\over\d V}}
\def\dW{{\d\over\d W}}
\def\dX{{\d\over\d X}}
\def\dY{{\d\over\d Y}}
\def\dZ{{\d\over\d Z}}

\def\upt{\p_t}
\def\upr{\p_r}
\def\upv{\p_v}
\def\upy{\p_y}
\def\upf{\p_f}
\def\upx{\p_x}
\def\upz{\p_z}
\def\up#1{\p_{#1}}

\def\pa#1#2{\f{\dsp \p#1}{\dsp \p#2}}
\def\pp#1#2{\f{\dsp \p^2 #1}{\dsp \p #2^2}}
\def\pam#1#2#3{\f{\dsp \p^2 #1}{\dsp \p #2 \p #3}}

\def\dm#1#2{\f{\dsp \d #1}{\dsp \d #2}}
\def\dd#1#2{\f{\dsp \d^2 #1}{\dsp \d #2^2}}

\def\dddot#1{\mathinner{\buildrel\vbox{\kern5pt\hbox{...}}\over{#1}}}


\def\be{\begin{equation}}
\def\ee{\end{equation}}
\def\bq{\begin{eqnarray}}
\def\eq{\end{eqnarray}}
\def\beq{\begin{eqnarray*}}
\def\eeq{\end{eqnarray*}}
\def\ext#1#2{G^{[#1]} #2_{|_{\!\!|_{#2=3D0}}} =3D 0}
\def\exta#1{G^{[#1]}}
\def\extb#1#2{G^{[#1]} #2_{|_{\!\!|_{#2=3D0}}} }
\def\enter{$\longleftarrow\!\!\!^|$}
\def\heading#1{\begin{center}{\LARGE #1} \end{center} \vspace{10mm}}

\def\de{\d ifferential equation}
\def\des{\d ifferential equations}
\def\fode{first order ordinary differential equation}
\def\fodes{first order ordinary differential equations}
\def\pde{partial differential equation}
\def\pdes{partial differential equations}
\def\sodes{second order ordinary differential equations}
\def\sode{second order ordinary differential equation}
\def\odes{ordinary differential equations}
\def\ode{ordinary differential equation}
\def\todes{third order ordinary differential equations}
\def\tode{third order ordinary differential equation}
\def\node{$n$th order ordinary differential equation}
\def\nodes{$n$th order ordinary differential equations}


\def\ie{{\it ie }}
\def\viz{{\it viz }}
\def\etal{{\it et al }}
\def\n{\nonumber}
\def\({\left (}
\def\){\right )}
\def\bi{\begin{itemize}}
\def\ei{\end{itemize}}
\def\z{&=&}
\def\lb{\left[ }
\def\rb{\right] }
\def\pic#1#2#3{\begin{center}\input #1.tex #2 =
\end{center}\begin{center}{\it #3}\end{center}}
\def\lrl{Laplace-Runge-Lenz}
\def\lrlv{Laplace-Runge-Lenz vector}
\def\kp{Kepler Problem}
\def\kdp{Kepler-Dirac Problem}
\def\noe{Noether's Theorem}
\def\re#1{(\ref{#1})}


\title{The harmony in the Kepler and related problems}
\author{M C
Nucci$^{\dagger}$ and P G L Leach$^{\ddagger}$\footnote{permanent
address: School of Mathematical and Statistical Sciences,
University of Natal, Durban 4041, Republic of South Africa}}
\date{$^{\dagger}$Dipartimento di Matematica e Informatica,
Universit\`a di Perugia\newline 06123 Perugia,
Italia\\$^{\ddagger}$GEODYSYC, Department of Mathematics,
University of the Aegean\newline Karlovassi 83 200, Greece}

\maketitle

\begin{abstract}
The technique of reduction of order developed by Nucci ({\it J
Math Phys} {\bf 37} (1996) 1772-1775) is used to produce nonlocal
symmetries additional to those reported by Krause ({\it J Math
Phys} {\bf 35} (1994) 5734-5748) in his study of the complete
symmetry group of the Kepler Problem.  The technique is shown to
be applicable to related problems containing a drag term which
have been used to model the motion of low altitude satellites in
the Earth's atmosphere and further generalisations.  A
consequence of the application of this technique is the
demonstration of the group theoretical relationship between the
simple harmonic oscillator and the Kepler and related problems.
\end{abstract}

\section{Introduction}

In a paper of a few years ago Krause \cite{Krause} introduced a new concept
into the study of the symmetries of ordinary differential equations.  He
called this a complete symmetry group and defined it by adding two properties
to the definition of a Lie symmetry group.  These were that the manifold of
solutions is an homogeneous space of the
group and the group is specific to the system, \ie no other system admits it.
This definition required the introduction of a new type of symmetry defined by
\begin{equation}
Y = \lb\int\xi (t,x_1,\ldots,x_N)\d t\rb\upt + \sum_{i = 1}^N\eta_i (t,x_1,\ldots,x_N)\up{x_i}. \label{1.1}
\end{equation}
This definition of a symmetry differs from that of a Lie point symmetry due to
the presence of the integral as the coefficient function of $\upt $.

As an illustration of the concept of a complete symmetry group
Krause used the Kepler problem and obtained three symmetries of
the type of \re{1.1}.  He claimed that these three symmetries
could not be obtained by means of the standard Lie point symmetry
analysis.  Naturally it was not long before this claim was shown
by Nucci \cite{Nucci} to be incorrect in the case of an autonomous
system.  In the case of the Kepler problem, an autonomous system,
one of the dependent variables can be taken to be the new
independent variable and the order of the system be reduced by
one.  An analysis of the reduced system for Lie point symmetries
leads to results different from the analysis of the original
system.  In particular the three additional nonlocal symmetries
obtained by Krause followed from point symmetries of the reduced
system.

One of the fundamental problems of mechanics is that of the Kepler
problem which describes the interaction of two point particles
with an inverse square law of attraction.  It is well-known that
this problem possesses the first integrals of the conservation of
the scalar energy, the vector of angular momentum and vectors in
the plane of the orbit known as Hamilton's vector \cite{Hamilton}
and the Laplace-Runge-Lenz vector
\cite{Ermanno,Herman,Bernoulli,Laplace,Runge,Lenz}.
The invariance Lie algebra of the first
integrals under the operation of taking the Poisson Bracket is
$so(4) $ (in the case of negative energy) and the Lie algebra of
the five Lie point symmetries of the equation of motion $A_2\oplus
so (3) $. The algebra of the complete symmetry group has not been
given. The elements of the five-dimensional algebra are
\begin{equation}
\begin{array}{ll}
X_1 = \upt &X_3 = x_2\up{x_3}-x_3\up{x_2} \n\\
X_2 = t\upt + \tth r\upr &X _4 = x_3\up{x_1}-x_1\up{x_3} \n\\
                              &X_5 = x_1\up{x_2}-x_2\up{x_1}.
\end{array} \label{1.2}
\end{equation}
The additional three nonlocal symmetries provided by Krause are
\begin{eqnarray}
Y_1 = 2\(\int x_1\d t\)\upt + x_1r\upr \n\\
Y_2 = 2\(\int x_2\d t\)\upt + x_2r\upr \n\\
Y_1 = 2\(\int x_3\d t\)\upt + x_3r\upr \label{1.3}
\end{eqnarray}
in which $r^2 = x_1^2+x_2^2+x_3^ 2.  $

There have been several other systems, generalisations of the
Kepler problem, which have been shown to have a similar set of
conserved quantities
\cite{Gorringe1,Gorringe2,Gorringe3,Gorringe4}.
Just as the Laplace-Runge-Lenz vector provides a
direct route to the equation of the orbit of the classical Kepler
problem, the corresponding vectors of the generalised Kepler
problems provide the same direct route to the equations of their
orbits. The Lie point symmetry associated with the
Laplace-Runge-Lenz vector of the classical Kepler problem is the
rescaling symmetry, $X_2 $.  The generalisations of the
Laplace-Runge-Lenz vector do not always have such an associated
Lie point symmetry.  In the case of the equation of motion
\begin{equation}
\ddot{\bfr} -\(\frac{\dot{g}}{2g} +\frac{3\dot{r}}{2r}\)\dot{\bfr} +\mu g\bfr = 0, \label{1.4}
\end{equation}
which is a variation of the model proposed by Danby
\cite {Danby,Mittelman1,Mittelman2,Leach1} for the
motion of a satellite in a low
altitude orbit subject to atmospheric drag, \viz
\begin{equation}
\ddot{\bfr} +\frac{\alpha\dot{\bfr}}{r^2} + \frac{\mu \bfr}{r^3} = 0 \label{1.5},
\end{equation}
and for which the generalisation of the Laplace-Runge-Lenz vector is
\begin{equation}
\bfJ = \frac{\dot{\bfr}\times\hat{\bfL}}{L} + \frac{\mu}{A^2}\hat{\bfr}, \label{1.6}
\end{equation}
where $L$ is the magnitude of the angular momentum, $\bfL$, and $A $ is a
constant of the motion defined through
\begin{equation}
L = A\(gr^3\)^{\ha}, \label{1.7}
\end{equation}
Pillay \etal\ \cite{Pillay} showed that, instead of the Lie point
symmetry for the classical Kepler problem, $X_2 $, the nonlocal
symmetry
\begin{equation}
G = -\ha\lb\int\frac{g'r}{g}\d t\rb\upt +r\upr \label{1.8}
\end{equation}
was the corresponding associated Lie symmetry.

In this paper we intend to demonstrate the
existence of far more nonlocal
symmetries for the classical Kepler problem than were reported by Krause.
We derive them as Lie point symmetries of a reduced system using the method of
Nucci \cite{Nucci1}.  By deriving these additional nonlocal symmetries in this
way we are able to say something definite about the expanded symmetry group.
The essence of the method of Nucci is to reduce the order of the system by
using the symmetry, $X_1 $, which is a statement of the autonomy of the system.
We recall that in a reduction of order the symmetry $Z $ which does not have
the property $[X_1,Z] =\lambda X_1 $ becomes an exponential nonlocal
symmetry \cite{Feix}.  By an exponential nonlocal symmetry we mean one of the
form
\begin{equation}
G = \exp\lb\int f\d t\rb\(\tau\upt+\eta_i\up{x_i}\) \label{1.8a}
\end{equation}
in which without a knowledge of the solution of the differential
equation $f$ is not an exact derivative and $\tau$ and the
$\eta_i$ are functions only of $t$, the $x_i$ and their
derivatives.  In this case we have nonlocal symmetries becoming
local on the reduction of order. Consequently we know that the
symmetries of the reduced equation quite possibly have zero Lie
Bracket with $X_1 $ and this will enable us to construct the
algebra. Further we shall show that these considerations which are
applicable to the classical Kepler problem can be extended to the
generalisations such as the one in \re{1.4}.

In the next section we review the reduction procedure of Nucci
\cite{Nucci} and in Section 3 we apply it to the classical Kepler
problem and see that there is a certain delicacy in the choice of
the new independent variable. For the sake of simplicity we work
in two dimensions. In Section 4 we make some observations about
these symmetries, the route to further simplification and the
algebra.  In Section 5 we obtain the results for Danby problem
\cite{Danby}, in Section 6 those for the generalised problem
represented by \re{1.4}, in Section 7 the symmetries for another
generalisation in which the force is not only not central but is
also angle dependent and in Section 8 we present our conclusions
and make some pertinent observations about going to the full three
dimensions.

 \section{The method of reduction of order}

 Consider the system of $N $ second order ordinary differential equations given by
 \begin{equation}
 \ddot{x_i} = f_i(x,\dot{x}),\quad i = 1,N \label{2.1}
 \end{equation}
 in which $t $ is the independent variable and $x_i, i = 1,N $ the $N $
dependent variables.  These equations may be considered as
equations from Newtonian mechanics, which was Krause's approach,
but there is no necessity for
that to be the case.  There is also no necessity for the dependent
variables to represent cartesian coordinates. Indeed there is no
need for the system to be of the second order.  It just so happens
that many of the equations which arise in practice have their
origins in Newton's Second Law and so are second order equations.
There is no requirement that the system be autonomous. In the case
of a nonautonomous system we can apply the standard procedure of
introducing a new variable $x_{N+ 1} = t $ and an additional first
order equation $\dot{x}_{N+ 1} = 1 $ so that the system becomes
formally autonomous. In our discussion we confine our attention to
autonomous systems.  We reduce the system \re{2.1} to a $2N
$-dimensional first order system by means of the change of
variables
\be
 \begin{array}{ccc}
 w_1 = x_1&\qquad & w_{N+ 1} = \dot{x_1}\n\\
 w_2 = x_2 && w_{N+ 2} = \dot{x_2}\n\\
 \vdots &&\vdots \n\\
 w_{N- 1} = x_{N- 1} && w_{2N- 1} = \dot{x}_{N- 1 } \n\\
 w_N = x_N && w_{2N} = \dot{x}_N
 \end{array}                    \label{2.2}
\ee
 so that the system \re{2.1} becomes
 \begin{equation}
 \dot{w_i} = g_i (x,w),\quad i = 1,2N, \label{2.3}
 \end{equation}
 where $g_i = w_{N+i} $ for $i = 1,N $ and $g_i = f_i $ for $i = N+1,2N $.

In the first step of the reduction of the original system \re{2.1}
we simply follow the conventional method used to reduce a higher
order system to a first order system.  Any optimisation is
performed in the further selection of the final variables.  This
selection may be motivated by the existence of a known first
integral, such as angular momentum, or some specific symmetry in
the original system \re{2.1}.

 We choose one of the variables $w_i $ to be the new independent variable $y $.
For the purpose of the development here we can make the identification $w_N = y
$.  By taking the quotients of the first order equations of the remaining
members of the set \re{2.2} with (\ref{2.3}N) we obtain the $(2N- 1) $-dimensional system
 \begin{equation}
 \dm{w_i}{w_N} = \frac{g_i}{g_N} =\frac{g_i}{w_{2N}},\quad i = 1,\ldots,
N- 1,N+ 1,\ldots,2N. \label{2.4}
 \end{equation}
 We do not attempt to calculate the Lie point symmetries of the system \re{2.4}
because the Lie point symmetries of a first order system are
generalised symmetries and one has to impose some {\it Ansatz} on
the form of the symmetry. Rather we select $n \leq N- 1 $ of the
variables to be the new dependent variables and rewrite the system
\re{2.4} as a system of $n $ second order equations plus
$2(N-n)-1$ first order equations.  The selection of the new
dependent variables is dictated by a number of considerations.
The first and foremost is that we must be able to eliminate the
unwanted variables from the system \re{2.4}.  After this condition
has been satisfied we may look to seek variables which reflect
some symmetry of the system, for example an ignorable coordinate
such as the azimuthal angle in a central force problem.

After the symmetries have been calculated, they can now be
translated back to symmetries of the original system as follows.
Suppose that the symmetry in the original variables is given by
 \begin{equation}
 G = \tau\upt +\eta_i\up{x_i}. \label{2.5}
 \end{equation}
The symmetry $G $ is first extended and then rewritten in terms of
the new coordinates as follows
 \begin{eqnarray}
 G^{[1]} \z \tau\upt +\eta_i\up{x_i} + (\dot{\eta}_i-\dot{x}_i\dot{\tau})\up{\dot{x}_i} \n\\
 \z \tau\upt + \zeta_i \up{w_i} \n\\
 \z \sigma\upy + \xi_i\up{u_i}, \label{2.6}
 \end{eqnarray}
where in the first line the summation is from $1 $ to $N $, in the
second from $1 $ to $N- 1 $ and $N+ 1 $ to $2N $ and in the third
over the number of dependent variables $u_i $ (the number cannot
be fixed in advance without a knowledge of the specific system);
$\zeta_i = \eta_i $ for $i = 1,N- 1 $ and $\zeta_i =
\dot{\eta}_i-\dot{g_i}\dot{\tau} $ for $i =N+ 1,2N $; $\sigma
=\eta_N $; $\xi_i =\zeta_j\p u_i/\p w_j $.  The only way that
$\tau $ appears in the symmetries of the reduced system is through
its derivative with respect to time.  If the nonlocality in the
original system occurs as a simple integral in $\tau $ of a
function of the original dependent variables, $x_i $, this will be
passed to the reduced system as a function of the new variables.
When the point symmetries of the reduced system are computed, the
form which the symmetries take in the original system can be
determined from \re{2.6}.  Since $\tau $ is determined as its
derivative with respect to time, the symmetry of the original
system must necessarily be nonlocal unless the derivative is an
exact differential.  We note in passing that there is no inherent
restriction on the nature of the symmetries. They could equally be
contact or generalised symmetries and the same considerations
would apply.  The only requirement is that $\tau $ be a simple
integral, not that the integrand be a point function.  However,
for the purposes of this paper we confine our attention to point
symmetries and the integrand in $\tau $ to a point function.

In addition to the nonlocal symmetries which may be collected by
this procedure the reduced system will have as point symmetries
those symmetries of the original system which have the correct Lie
bracket with the symmetry $\upt $ which is at the basis of the
reduction of order outlined above.  Thus, if $G $ is a symmetry of
the original system and
 \begin{equation}
 [G,\upt] = \lambda\upt, \label{2.6a}
 \end{equation}
$G $ will, when expressed in the appropriate coordinates, be a
point symmetry of the reduced system whereas, if
 \begin{equation}
 [G,\upt] \neq \lambda\upt, \label{2.7}
 \end{equation}
$G $ will not be a point symmetry of the original system but an
exponential nonlocal symmetry \cite{Feix}.  Consequently there is
the potential for a loss of symmetry in the reduction process just
as there is the hope of an increase in the total number of
symmetries, both point and nonlocal, known for the original
system.

 \section{Lie point symmetries of the reduced Kepler problem}

The Lagrangian for the two-dimensional Kepler problem is
\begin{equation}
L = \ha\(\dot{r}^2+r^2\dot{\theta}^2\) +\frac{\mu}{r} \label{3.1}
\end{equation}
in plane polar coordinates and the two equations of motion are
\begin{eqnarray}
\ddot{r} -r\dot{\theta}^2 \z -\frac{\mu}{r^2} \label{3.2}\\
r\ddot{\theta} + 2\dot{r}\dot{\theta} \z 0. \label{3.3}
\end{eqnarray}
We introduce the new variables and their time derivatives
\begin{equation}
\begin{array}{lcl}
w_1 = r &\qquad&\dot{w}_1 = w_3\\
w_2 = \theta &&\dot{w}_2 = w_4\\
w_3 =\dot{r} &&\dot{w}_3 = w_1w_4^2-{\dsp\frac{\mu}{w_1^2}}\\
w_4 = \dot{\theta} &&\dot{w}_4 = -{\dsp\frac{2w_3w_4}{w_1}}.
\end{array} \label{3.4}
\end{equation}
In accordance with the development in the previous section we select $w_2 $ to
be the new independent variable $y $.  The left side of \re{3.4} leads to the
reduced system
\begin{eqnarray}
\dm{w_1}{y} \z \frac{w_3}{w_4} \label{3.5}\\
\dm{w_3}{y} \z w_1w_4- \frac{\mu}{w_1^2w_4} \label{3.6}\\
\dm{w_4}{y} \z -\frac{2w_3}{w_1}. \label{3.7}
\end{eqnarray}
>From \re{3.5} we have $w_3 = w_4w'_1 $, where the prime denotes
differentiation with respect to the new independent variable, $y
$, and we replace \re{3.7} by
\begin{equation}
\dm{w_4}{y} = -\frac{2w_4w'_1}{w_1}. \label{3.8}
\end{equation}
In \re{3.6} we replace $w_3 $ by $w_4w'_1 $ to obtain
\begin{equation}
w_4w''_1-\frac{w_4w'{}_1^2}{w_1} = w_1w_4-\frac{\mu}{w_1^2w_4}. \label{3.9}
\end{equation}

By our replacement of $w_3 $ we have not precisely decided that
the variables $u_1 $ and $u_2 $ are to be $w_1 $ and $w_4 $.  If
we do make this identification, we obtain a system of two
equations, one of the second order and one of the first order,
\viz
\begin{eqnarray}
u''_1 \z 2\frac{u'{}_1^2}{u_1} + u_1-\frac{\mu}{u_1^2u_2^2}\label{3.10}\\
u'_2 \z - 2\frac{u'_1u_2}{u_1}.\label{3.11}
\end{eqnarray}
We observe that \re{3.11} is trivially integrated to give
$u_1^2u_2 $ is a constant.  (This is a consequence, naturally, of
the symmetry $\up{\theta} $ of the original system \re{3.2} and
\re{3.3} which is a reflection of the fact that $\theta $ is an
ignorable coordinate.) Consequently we may just as well define our
new variables to be
\begin{equation}
u_1 = w_1\quad\mbox{\rm and}\quad \tilde{u}_2 = u_1^2u_2\label{3.12}
\end{equation}
so that the system of equations we are to consider is
\begin{eqnarray}
u''_1 \z 2\frac{u'{}_1^2}{u_1} + u_1-\frac{\mu u_1^2}{\tilde{ u}_2^2}\n\\
\tilde{u} '_2 \z 0.\label{3.14}
\end{eqnarray}
Hereafter we drop the tilde.

We calculate the Lie point symmetries of the system \re{3.14} using the
well-known interactive program developed by Nucci \cite{Nucci1} and obtain the symmetries
\begin{eqnarray}
X_1 \z \upy\n\\
X_2 \z 2u_1\up{u_1} + u_2\up{u_2}\n\\
X_3 \z u_1\(\mu u_1-u_2^2\)\up{u_1}\n\\
X_4 \z u_1^2\cos y\up{u_1}\n\\
X_5 \z -u_1^2\sin y\up{u_1}\n\\
X_6 \z \(\mu u_1-u_2^2\)\cos y\upy +u_1\(2\mu u_1-u_2^2\)\sin y\up{u_1}\n\\
X_7 \z -\(\mu u_1-u_2^2\)\sin y\upy +u_1\(2\mu u_1-u_2^2\)\cos y\up{u_1}\n\\
X_8 \z u_2^2\cos 2y\upy -u_1\(\mu u_1-u_2^2\)\sin 2y\up{u_1}\n\\
X_9 \z u_2^2\sin 2y\upy + u_1\(\mu u_1-u_2^2\)\cos 2y\up{u_1}.\label{3.15}
\end{eqnarray}
Some of these symmetries are readily identified, but not many of
them.  Clearly $X_1 $ represents the rotational invariance of the
system and constitutes the subalgebra, $so(2) $, of the original
system of equations. In $X_2 $ we recognise the rescaling symmetry
closely associated with the Laplace-Runge-Lenz vector.

The other symmetries are not so easy to identify without making
some calculation. To obtain the form of the symmetries in the
original coordinates we must make use of \re{2.6}.  If we write
the symmetry in the original coordinates as
\begin{equation}
G = \tau\upt +\eta\upr +\zeta\up{\theta},\label{3.16}
\end{equation}
in terms of the new variables, the symmetry has the form
\begin{equation}
\tilde{G} = \zeta\upy +\eta\up{u_1} + \lb\dot{\zeta} +
\(\frac{2\eta}{r} -\dot{\tau}\)\dot{\theta}\rb r^2\up{u_2},
\label{3.17}\end{equation} where the coefficient functions are all
expressed in terms of the original variables. The calculations are
not particularly interesting and we simply list the symmetries.
They are
\begin{eqnarray}
X_1 \z \up{\theta}\n\\
X_2 \z 3t\upt + 2r\upr\n\\
X_3 \z 2\lb\mu\int r^2\d t-L^2t\rb\upt +\mu r\(r^2-L^2\)\upr\n\\
X_4 \z 2\int r\cos\theta\d t\upt +r^2\cos\theta\upr\n\\
X_5 \z 2\int r\sin\theta\d t\upt +r^2\sin\theta\upr\n\\
X_6 \z \int\lb\(3\mu r-L^2\)\sin\theta +\mu\frac{\dot{r}}{\dot{\theta}}\cos\theta\rb\d
t\upt +r\sin \theta\(2\mu r-L^2\)\upr \n\\
&&+\cos\theta\(\mu r-L^2\)\up{\theta}\n\\
X_7 \z \int\lb\(3\mu r-L^2\)\cos\theta
-\mu\frac{\dot{r}}{\dot{\theta}}\sin\theta\rb\d t\upt+r\cos
\theta\(2\mu r-L^2\)\upr\n\\
&& -\sin\theta\(\mu r-L^2\)\up{\theta}\n\\
X_8 \z 2\mu\int r\sin 2\theta\d t\upt +r\(\mu r-L^2\)\sin
2\theta\upr -L^2\cos 2\theta\up{\theta}\n\\ X_9 \z 2\mu\int r\cos
2\theta\d t\upt +r\(\mu r-L^2\)\cos 2\theta\upr +L^2\sin
2\theta\up{\theta} \label{3.18}
\end{eqnarray}
so that we see that we lost only the time translation symmetry in the reduction of order
 and so gained six additional symmetries.

\section{The Lie point symmetries of the reduced Kepler problem: further
considerations}

Further interpretation of the Lie point symmetries of the reduced
generalised Kepler problem is facilitated by the redefinition of
the symmetries given in \re{3.15}.  We do not alter the
definitions of the first three symmetries, but we shall include
them in this new listing for the sake of completeness.  We now
have the set of symmetries
\begin{eqnarray}
X_1 \z \upy\n\\
X_2 \z 2u_1\up{u_1} + u_2\up{u_2}\n\\
X_3 \z u_1\(\mu u_1-u_2^2\)\up{u_1}\n\\
X_{4\pm} \z X_4\pm iX_5 = \e^{\pm i y}u_1^2\up{u_1}\n\\
X_{5\pm} \z X_6\mp i X_7 =
 \e^{\pm i y}\lb \(\mu u_1-u_2^2\)\upy \mp u_1\(2\mu u_1-u_2^2\)\up{u_1}\rb \n\\
X_{6\pm} \z X_8\pm iX_9 = \e^{\pm 2i y}\lb u_2^2\upy \pm u_1\(\mu
u_1-u_2^2\)\up{u_1}\rb.\label{4.1}
\end{eqnarray}
We may search for the first integrals/invariants associated with
each of these symmetries in the usual way. By way of concrete
example we take $X_{4+} $.  The invariants of the first extension
of $X_{4+} $, \viz
\begin{equation}
X_{4+}^{[1]} = \e^{iy}\lb u_1^2\up{u_1} +\(2u_1u'_1+iu_1^2\)\up{u'_1}\rb,\label{4.2}
\end{equation}
are found from the associated Lagrange's system
\begin{equation}
\frac{\d y}{0} =\frac{\d u_1}{u_1^2} =\frac{\d u_2}{0}
 =\frac{\d u'_1}{2u_1u'_1+iu_1^2}\label{4.3}
\end{equation}
and are
\begin{equation}
\alpha = y,\quad \beta = u_2\quad\mbox{\rm and}\quad \gamma =\frac{u'_1+iu_1}{u_1^2},\label{4.4}
\end{equation}
where the first two are by inspection and the third comes from the
solution of the second and fourth of \re{4.3}.  The
integral/invariant is a function of these three arguments and is
found by demanding that the total derivative with respect to $y $
be zero when the differential equations \re{3.14} are taken into
account.  We obtain the associated Lagrange's system
\begin{equation}
\frac{\d\alpha}{1} =\frac{\d\beta}{0} =\frac{\d\gamma}{-i\gamma
-\dsp{\frac{\mu}{\beta^2}}}\label{4.5}
\end{equation}
which gives $\beta $ as one of the characteristics and
\begin{equation}
\omega_+ = \e^{iy}\(\frac{u'_1+iu_1}{u_1^2}-\frac{i\mu}{u_2^2}\)\label{4.6}
\end{equation}
as the second characteristic. Both $\beta $ and $\omega $ are
first integrals/invariants of the system \re{3.14}.  Naturally we
recognise the former as the angular momentum.

If we perform the same calculation with $X_{4-} $, we obtain a similar result, \viz
\begin{equation}
\omega_- = \e^{- iy}\(\frac{u'_1-iu_1}{u_1^2} +\frac{i\mu}{u_2^2}\).\label{4.7}
\end{equation}
The two can be combined into one convenient expression given by
\begin{equation}
J_{\pm} = \e^{\pm iy}\(v'\mp iv\),\quad v = \mu -\frac{u_2^2}{u_1},\label{4.8}
\end{equation}
where we have made use of the constancy of $u_2 $ to write
$J_{\pm} = u_2^2\omega_{\pm} $. For the system \re{3.14} $J_{\pm}
$ are two invariants and for the original system, \re{3.1} and
\re{3.2}, the two components of the first integral known as the
Laplace-Runge-Lenz vector, written in complex form.

In \re{4.8} we introduced a new variable $v $.  If we use this variable instead
of $u_1 $ in the system \re{3.14}, we obtain the system
\begin{eqnarray}
v''+v \z 0\n\\
u'_2 \z 0\label{4.9}
\end{eqnarray}
so that the natural variable which arises from the invariants of
$X_{4\pm} $ is a variable which further simplifies the reduced
system.  This is an interesting phenomenon for we are obtaining
natural variables as we progress through the process of
determining the symmetries of the reduced equation.  It behooves
us to rewrite the symmetries in terms of this new variable.  We
find that
\begin{eqnarray}
X_1 \z \up{u_2}\n\\
X_2 \z \upy\n\\
X_3 \z v\upv\quad\mod(u_2^2)\n\\
X_{4\pm} \z  \e^{\pm i y}\upv\quad\mod(u_2^2)\n\\
X_{6\pm} \z  \e^{\pm 2i y}\lb \upy \pm iv\upv\rb\quad\mod(u_2^2)\n\\
X_{8\pm} \z  \e^{\pm i y}\lb v\upy \pm i v^2\upv\rb\quad\mod(u_2^2).\label{4.101}
\end{eqnarray}
which is certainly a simpler appearance.

In the simpler form presented in \re{4.11} we see that the
calculation of the integrals/invariants is simpler.  For example,
if we take $X_{6\pm} $, the first set of characteristics comes
from the solutions of the associated Lagrange's system
\begin{equation}
\frac{\d y}{1} =\frac{\d v}{iv} =\frac{\d v'}{-iv'- 2v}.\label{4.11}
\end{equation}
The characteristics are
\begin{equation}
\alpha = v\e^{\mp iy}\quad\mbox{\rm and}\quad \beta = vv'\mp iv^2.\label{4.12}
\end{equation}
The condition for the function to be an invariant (in this case not a first
integral of the reduced system since the independent variable is explicitly present)
 is that these two characteristics satisfy the first order equation
\begin{equation}
\frac{\d\alpha}{\alpha} =\frac{\d\beta}{\beta},\label{4.13}
\end{equation}
whence the invariants are given by
\begin{equation}
I_{\pm} =\frac{\beta}{\alpha} = (v'\mp iv)\e^{\pm iy}\label{4.14}
\end{equation}
which are, of course, the two components of the Laplace-Runge-Lenz vector.  We
note that there is also a first integral associated with each of $X_{6\pm} $
and this is the angular momentum which, in these coordinates, is an ignorable coordinate.

For the sake of completeness we list the first
integrals/invariants associated with the symmetries listed in
\re{4.101}. They are given in the same order and with the same
subscripts as the symmetries are listed.
\begin{equation}
\begin{array}{lclcl}
G_1 &\quad& I_1 = L &\quad&I_2 = J_+J_- = 2L^2E+\mu^2\\
G_2 && I_1 = L && I_2 = \dsp{\frac{J_+}{J_-}}\\
G_3 && I_1 = L && I_2 = \dsp{\frac{J_+}{J_-}}\\
G_{4\pm} && I_1 = L && I_{2\pm} = J_{\pm}\\
G_{5\pm} && I_1 = L && I_{2\pm} = \dsp{\frac{J_{\pm}}{J_{\mp}}}\\
G_{6\pm} && I_1 = L && I_{2\pm} = J_{\pm}.
\end{array}\label{4.15}
\end{equation}
We note that in the second integral associated with $G_1 $ we have expanded
the product into the standard expression relating the square of the magnitude
of the Laplace-Runge-Lenz vector with the energy and the angular momentum.
For the second integral of $G_{5\pm} $ we could have equally written
$I_2 = J_+/J_- $, but we chose to maintain the pattern of $\pm $.

\section{The Kepler problem with drag}

In the introduction we referred to the model proposed by Danby
\cite{Danby} for the motion of a low altitude satellite subjected
to a resistive force due to the Earth's atmosphere described by
the equation of motion
\begin{equation}
\ddot{\bfr} +\frac{\alpha\dot{\bfr}}{r^2} + \frac{\mu \bfr}{r^3} = 0 \label{10.1},
\end{equation}
where $\alpha $ and $\mu $ are constants.  Since the direction of the angular
momentum is a constant, we may analyse the problem in two dimensions
 using plane polar coordinates, $(r,\theta) $.  The two equations of motion are
\begin{eqnarray}
\ddot{r} -r\dot{\theta}^2 +\frac{\alpha\dot{r}}{r^2} + \frac{\mu}{r^2} \z 0 \label{10.2}\\
r\ddot{\theta} + 2\dot{r}\dot{\theta} +\frac{\alpha\dot{\theta}}{r}\z 0. \label{10.3}
\end{eqnarray}
We introduce the new variables and their time derivatives
\begin{equation}
\begin{array}{lcl}
w_1 = r &\qquad&\dot{w}_1 = w_3\\
w_2 = \theta &&\dot{w}_2 = w_4\\
w_3 =\dot{r} &&\dot{w}_3 = w_1w_4^2-\dsp{\frac{\alpha w_3}{w_1^2}} -{\dsp\frac{\mu}{w_1^2}}\\
w_4 = \dot{\theta} &&\dot{w}_4 = -{\dsp\frac{2w_3w_4}{w_1}}
-\dsp{\frac{\alpha w_4}{w_1^2}}.
\end{array} \label{10.4}
\end{equation}
We again select $w_2 $ to
be the new independent variable $y $.  The right side of \re{10.4} becomes
\begin{eqnarray}
\dm{w_1}{y} \z \frac{w_3}{w_4} \label{10.5}\\
\dm{w_3}{y} \z w_1w_4-\frac{\alpha w_3}{w_1^2w_4}- \frac{\mu}{w_1^2w_4} \label{10.6}\\
\dm{w_4}{y} \z -\frac{2w_3}{w_1}-\frac{\alpha w_4}{w_1^2}. \label{10.7a}
\end{eqnarray}

In this case the choice of \re{10.5} to eliminate $w_3 $ is obvious and we
 obtain the two-dimensional system
\begin{eqnarray}
w_1''w_4+w'_1w'_4 \z w_1w_4-\frac{\alpha w'_1}{w_1^2} -\frac{\mu}{w_1^2w_4}
\label{10.7}\\
w'_4 \z -\frac{2w'_1w_4}{w_1} -\frac{\alpha}{w_1^2}.\label{10.8}
\end{eqnarray}
In the case of \re{10.8} we can easily manipulate it to obtain
\begin{equation}
\(w_1^2w_4\)'=\alpha\quad\Leftrightarrow\quad w_1^2w_4 = -\alpha y+\beta,\label{10.9}
\end{equation}
where $\beta $ is a constant of integration, which indicates that the angular
momentum is not conserved.  We have a choice of defining the new variable
$u_2 $ as either $w_1^2w_4 $, which is not conserved but is a convenient
variable for manipulations, or $w_1^2w_4+\alpha y $, which is conserved
but is not a convenient variable for manipulation.  For the present we make
the former choice. In equation \re{10.7}
we make use of \re{10.9} to eliminate $w_4 $ and $w'_4 $ to obtain
\begin{equation}
\(-\frac{1}{w_1}\)''(\alpha y+\beta)^2+ 2\alpha\(-\frac{1}{w_1}\)'(\alpha y+\beta)
+\(-\frac{1}{w_1}\) (\alpha y+\beta)^2+\mu = 0.\label{10.10}
\end{equation}
We introduce the second new variable as
\begin{equation}
u_1 = -\frac{\alpha y+\beta}{w_1}\label{10.11}
\end{equation}
so that \re{10.10} becomes
\begin{equation}
u''_1+u_1 = -\frac{\mu}{\alpha y+\beta}.\label{10.12}
\end{equation}
Clearly we could regain the equation of a simple harmonic oscillator
 by means of the further change of variable $v =u_1-u_{1ps} $, where $u_{1ps} $
is a particular solution of \re{10.12}. Consequently the Kepler
problem with drag differs from the standard Kepler problem in that
the second equation of the reduced system, \viz
\begin{equation}
u'_2 = -\alpha,\label{10.13}
\end{equation}
has a nonzero right side. This could be eliminated by now making
the second choice mentioned above.  Consequently we make a final
change of variables
\begin{eqnarray}
v_1 \z u_1-\int\frac{\mu\sin (y-s )\d s}{\alpha s-\beta}\n\\
v_2 \z u_2+\alpha\label{10.14}
\end{eqnarray}
to obtain the same reduced system, \re{4.9},
as we had for the Kepler problem when the introduced the sensible coordinates.
 Naturally we obtain the same set of symmetries as given in \re{4.101}.

The Lie point symmetries of the reduced system
translate into the symmetries
\begin{eqnarray}
\Gamma_1 \z \(\int\frac{\d t}{r^2\dot{\theta}}\)\upt\n\\
\Gamma_2 \z 2\lb\int\frac{(\alpha r+I')\d t}{r (\alpha\theta -\beta)}\rb\upt
+\lb\frac{\alpha r+I'}{\alpha\theta -\beta}\rb\upr -\up{\theta}\n\\
\Gamma_3 \z \lb 2t+\int\frac{I\d t}{r (\alpha\theta -\beta)}\rb\upt +\(r+\frac{I}{\alpha\theta
-\beta}\)\upr\n\\
\Gamma_{4\pm} \z 2\lb\int\frac{\e^{\pm i\theta}\d t}{r (\alpha\theta -\beta)}\rb\upt
+\lb\frac{\e^{\pm i\theta}}{\alpha\theta -\beta}\rb\upr\n\\
\Gamma_{5\pm} \z 2\left\{\int\e^{\pm i\theta}\lb\frac{\alpha r+I}{r
(\alpha\theta -\beta)}
\mp\(2+\frac{I}{r (\alpha\theta -\beta)}\)\rb\d t\right\}\upt\n\\&&
+\e^{\pm 2i\theta}\lb\frac{\alpha r+I'}{\alpha\theta -\beta}
\mp i\(r+\frac{I}{\alpha\theta -\beta}\)\rb\upr -\e^{\pm 2i\theta}\up{\theta}\n\\
\Gamma_{6\pm} \z \left\{\int\e^{\pm i\theta}\lb\(r (\alpha\theta -\beta) +I\)
\(\frac{\alpha r+I'}{r (\alpha\theta -\beta)} \mp i\(3+\frac{2I}{r
(\alpha\theta -\beta)}\)\)\right.\right.\n\\&&\left.\left.
 +\alpha r+I'+\frac{\dot{r}}{\dot{\theta}} (\alpha\theta -\beta)
\rb\d t\right\}\upt +\e^{\pm i\theta}\(r (\alpha\theta -\beta)
+I\)\lb\frac{\alpha r+I'}{\alpha\beta -\beta }\right.\n\\&&\left.
\mp i\(r+\frac{I}{\alpha\theta -\beta}\)\rb\upr -\e^{\pm i\theta}\lb r
(\alpha\theta -\beta) +I\rb\up{\theta},
\label{10.15}
\end{eqnarray}
where $I$ stands for the integral introduced in (\ref{10.14}a) and $I'$ its
derivative with respect to $\theta$, for the original system, \re{10.1}.

In addition to the symmetries listed in \re{10.15} equation \re{10.1}
 has the point symmetry $\upt $ which was the symmetry used for the reduction of order.
Consequently we can conclude that algebraically the Kepler problem
and the Kepler problem with drag are identical.

In the above derivation we have followed a line of development in which
observation and experience play major roles in reducing the system \re{10.4} to
the simplest possible form.  The need for both are considerably obviated when
the interactive Lie symmetry solver devised by Nucci \cite{Nucci1} is used.
The equations to be solved suggest the appropriate variables since they are the
characteristics of the partial differential equations to be solved.  We
illustrate this in the case of the variable related to angular momentum with
the following tableau
$$w_2=y$$
$$w_3 = {{\mbox d} w_1\over {\mbox d} y}w_4$$
$w_4=u_1,w_1=u_2$
$$\dot u_1=-2{\dot u_2 u_1 \over u_2}-{\alpha \over u_2^2}$$
$$\ddot u_2={(u_2^2 + 2 \dot u_2^2) u_1^2 u_2 - \mu \over u_1^2 u_2^2}$$

$$w_4={w_5\over w_1^2}$$
$w_5=u_1,w_1=u_2$
$$\dot u_1=-\alpha$$
$$\ddot u_2={ - \mu u_2^3 + u_1^2 u_2^2 + 2 u_1^2 \dot u_2^2\over u_1^2 u_2}$$

$$w_5=w_6-\alpha y$$
$w_6=u_1,w_1=u_2$
$$\dot u_1=0$$
$$\ddot u_2={(\alpha y - u_1)^2 (u_2^2 + 2 \dot u_2^2) - \mu u_2^3\over
 (\alpha y - u_1)^2 u_2}.$$
(We have used $w_4$, $w_5$ and $w_6$ to successively define a new variable
which is a candidate for selection as $u_1$.  We are not introducing additional
variables.)

\section{The generalisation of the Kepler problem with drag}

Equation \re{1.4} has, in two dimensions, the two components of the equation of motion
\begin{eqnarray}
\ddot{r} \z r\dot{\theta}^2+\ha\(\frac{g'}{g} +\frac{3}{r}\)\dot{r}^2-\mu gr\n\\
\ddot{\theta} \z \frac{\dot{r}\dot{\theta}}{2r}\(\frac{g'}{g} -\frac{1}{r}\).\label{5.1}
\end{eqnarray}
We introduce the variables $w_i,i = 1, 4 $ as above.  Now the system of first
order equations in these variables is
\begin{eqnarray}
\dot{w}_1 \z w_3\n\\
\dot{w}_2 \z w_4\n\\
\dot{w}_3 \z w_1w_4^2+\ha\(\frac{g'}{g} +\frac{3}{r}\)w_3^2-\mu gw_1\n\\
\dot{w}_4 \z \frac{w_3w_4}{2w_1}\(\frac{g'}{g} -\frac{1}{w_1}\).\label{5.2}
\end{eqnarray}

As the new independent variable we take again $y =w_2 $.  The system \re{5.2} becomes
\begin{eqnarray}
\dm{w_1}{y} \z \frac{w_3}{w_4}\quad\Leftrightarrow\quad w_3 = w_4w_1'\n\\
\dm{w_3}{y} \z w_1w_4+\ha\(\frac{g'}{g} +\frac{3}{r}\)w_4w'{}_1^2-\mu g\frac{w_1}{w_4}\n\\
\dm{w_4}{y} \z \frac{w_3}{2w_1}\(\frac{g'}{g}w_1-1\).\label{5.3}
\end{eqnarray}
When we substitute for $w_3 $, the third of \re{5.3} is easily integrated to give
\begin{equation}
A =\(\frac{w_1}{g}\)^{\ha}w_4,\label{5.4}
\end{equation}
where $A $ is an arbitrary constant of integration. The right-hand
side is the same function as the characteristic of the parabolic
partial differential equation produced when the system \re{5.3} is
analysed using the code developed by Nucci \cite{Nucci1} and this
is an appropriate choice for one of the variables.  We take $u_1
=w_1 =r $ and $u_2 =\dot{\theta} (r/g)^{1/2} $ so that with the
elimination of $w_3 $ from \re{5.3} we have, after a certain amount
of simplification, the system of two equations
\begin{eqnarray}
u_2u_1'' \z u_1u_2+ 2\frac{u_2u'{}_1^2}{u_1} -\frac{\mu u_1^2}{u_2}\n\\
u_2' \z 0.\label{5.5}
\end{eqnarray}
We observe that \re{5.5} is precisely the system \re{3.14} and so
we immediately introduce the new variable $v = \mu -u_2^2/u_1 $ to
obtain the simpler system \re{4.9} which has the symmetries listed
in \re{4.101}.  In terms of the original variables these
symmetries are
\begin{eqnarray}
X_1 \z \up{\theta} \n\\
X_2 \z -\(\int\frac{g'r}{g}\d t\)\upt + 2r\upr \n\\
X_3 \z \ha\lb\int\(\mu rg -r\dot{\theta}^2\)\(\frac{1}{r} -\frac{g'}{g}\)\d t\rb\upt
+\(\mu rg -r\dot{\theta}^2\)\upr\n\\
X_{4\pm} \z \ha\lb\int\e^{\pm i\theta}\(g-rg'\)\d t\rb\upt
+\lb\e^{\pm i\theta}rg\rb\upr\n\\
X_{5\pm} \z \ha\left\{\int\e^{\pm i\theta}\lb\frac{gr}{\dot{\theta}^2}
\(\mu -\frac{\dot{\theta}^2}{g}\)^2\(\frac{1}{r} -\frac{g'}{g}\)
\pm 2i\(\mu -\frac{\dot{\theta}^2}{g}\) +\frac{2\dot{r}\dot{\theta}}{gr}\rb\d
t\right\}\upt\n\\
&& + \left\{\e^{\pm i\theta}\frac{gr}{\dot{\theta} 2}\(\mu -\frac{\dot{\theta}^2}{g}\)^2\right\}
\upr + \left\{\e^{\pm i\theta}\(\mu -\frac{\dot{\theta}^2}{g}\)\right\}
\up{\theta}\n\\
X_{6\pm} \z \pm\ha i\left\{\int\e^{\pm 2i\theta}\(3 +\frac{rg'}{g} +\frac{\mu
 (g-rg')}{\dot{\theta}^2}\)\d t\right\}\upt\n\\&&
\pm i\left\{\e^{\pm 2i\theta}r\(\frac{\mu g}{\dot{\theta}^2} - 1\)\right\}\upr
+ \e^{\pm 2i\theta}\up{\theta}.
\label{5.6}
\end{eqnarray}

\section{An example with an angle-dependent force}

Sen \cite{Sen} obtained conserved quantities similar to those of the Kepler
 problem for the Hamiltonian
\begin{equation}
H =
\ha\(p_r^2+\frac{p_{\gt}}{r^2}\)-\frac{\mu}{r}-
\frac{\ga\sin\lb\ha(\gt-\beta)\rb}{r^{1/2}}
\end{equation}
in which the potential depends upon the azimuthal angle and $\mu$, $\ga$ and
$\beta$ are constants.  Subsequently Gorringe and Leach \cite{Gorringe5}
 showed that the equation of motion
\begin{equation}
\ddot{\bfr} +g\hat{\bfr} +h\hat{\bf\theta} = 0,\label{11.1}
\end{equation}
where
\begin{equation}
g =\frac{U''(\theta) +U (\theta)}{r^2} + 2\frac{V'(\theta)}{r^{3/2}}\quad\mbox{\rm and}
\quad h =
\frac{V (\theta)}{r^{3/2}},
\label{11.2}
\end{equation}
or
\begin{eqnarray}
\ddot{r} -r\dot{\theta}^2+g = 0\label{11.3}\\
r\ddot{\theta} + 2\dot{r} \dot{\theta} +h = 0\label{11.4}
\end{eqnarray}
in plane polar coordinates, could be solved for the orbit equation
in a manner similar to that of the Kepler problem since it also possessed a
Laplace-Runge-Lenz vector. The only restrictions on the functions $U (\theta) $
and $V (\theta) $ are that they be differentiable.

We make the same reduction as in the previous cases to arrive at the two-dimensional system
\begin{eqnarray}
w_1w_1''w_4^2-2w_1'{}^2w_4^2-w_1^2w_4^2 =w_1'h-gw_1\label{11.5}\\
w_1w_4w_4'+ 2w_1'w_4^2 = -h,\label{11.6}
\end{eqnarray}
where $w_1 =r $ and $w_4 =\dot{\theta} $ as before.

The Laplace-Runge-Lenz vector for equation \re{11.1} is \cite{Gorringe5}
\begin{equation}
\bfJ =\dot{\bfr}\times\bfL -U\hat{\bfr} -\lb U'+
2r^{1/2}V\rb\hat{\bf\theta},\label{11.6a}
\end{equation}
where $\bfL :=\bfr\times\dot{\bfr} $ is the angular momentum. If
we take the two cartesian components of $\bfJ $, \viz $J_x $ and
$J_y $, and combine them we obtain
\begin{eqnarray}
J_{\pm} \z - J_x\pm iJ_y\n\\
\z \lb \(r^3\dot{\theta}^2-U\)\pm i\(-r^2\dot{r}\dot{\theta} -U'- 2r^{1/2}V\)
\rb\e^{\pm i\theta}\n\\
\z \lb\(w_1^3w_4^2-U\)\pm i\(-w_1^2w_4w_3-U'- 2w_1^{1/2}V\)\rb\e^{\pm iy}\n\\
\z \lb\(\frac{L^2}{w_1} -U\)\pm i\(\frac{L^2}{w_1} -U\)'\rb\e^{\pm iy}.\label{11.7}
\end{eqnarray}
We see that, when we write the components of the
Laplace-Runge-Lenz vector in this form, we have the same structure
as for the standard Kepler problem.  (One could call the
components the Ermanno-Bernoulli constants in honour of the
original discoverers of these conserved quantities.) Immediately
we have the clue to the identification of one of the new variables
and we let
\begin{equation}
u_1 =w_1^3w_4^2-U =\frac{L^2}{w_1} -U\label{11.8}
\end{equation}
so that the Ermanno-Bernoulli constants for \re{11.1} are
\begin{equation}
J_{\pm} =\(u_1\pm iu_1'\)\e^{\pm iy}.\label{11.9}
\end{equation}
The identification of the second variable is more delicate.
Equation \re{11.6} can be written in terms of the magnitude of the
angular momentum, $L $, as
\begin{equation}
LL'= -w_1^{3/2}V (y)\label{11.13}
\end{equation}
and, when \re{11.8} is taken into account, this becomes
\begin{equation}
0 =\frac{L'}{L^2} +\frac{V (y) }{(u_1+U (y))^{3/2}}.\label{11.14}
\end{equation}
>From \re{11.9} we have
\begin{eqnarray}
u_1 \z \ha\(J_+\e^{-iy} +J_-\e^{+iy}\)\n\\
\z J\cos y\label{11.15}
\end{eqnarray}
since $J_- =J_+^*$ and we have written $J =|J_+| =|J_-| $.
 Then we can use \re{11.14} to define a new variable
\begin{equation}
u_2 =\frac{1}{L} -\int\frac{V (y) \d y}{(J\cos y+U (y))^{3/2}}.\label{11.16}
\end{equation}

The reduced system of equations is
\begin{eqnarray}
u_1''+u_1 \z 0\n\\
u_2' \z 0\label{11.10}
\end{eqnarray}
which is just the reduced system we obtained for the standard Kepler problem
and so it has the symmetries given in \re{4.101}.  These
translate to
\begin{eqnarray}
\Gamma_1 \z 3\(\int r^2\dot{\theta}\d t \)\upt + 2r^3\dot{\theta}\upr\n\\
\Gamma_2 \z \lb\int\(\frac{2U'}{r^2\dot{\theta}^2}
+\frac{3Vr^2\dot{\theta}}{U+J\cos\theta}\)\d t\rb\upt
+\lb\frac{U'}{r^2\dot{\theta}^2} +\frac{2Vr^3\dot{\theta}}{U+J\cos\theta}\rb\upr
-\up{\theta}\n\\
\Gamma_3 \z 2\lb t-\int\frac{U\d t}{r^3\dot{\theta}^2}\rb\upt
+\lb r-\frac{U}{r^2\dot{\theta}^2}\rb\upr\n\\
\Gamma_{4\pm} \z 2\(\int\frac{\e^{\pm i\theta}}{r^3\dot{\theta}^2}\d t\)\upt +
\frac{\e^{\pm i\theta}}{r^2\dot{\theta}^2}\upr\n\\
\Gamma_{5\pm} \z \left\{\int\e^{\pm 2i\theta}\lb\frac{3Vr^2\dot{\theta}}{U
+J\cos\theta} +
\frac{2 (U' \mp iU)}{r^3\dot{\theta}^2}\rb\d t\right\}\upt\n\\& &
-\e^{\pm 2i\theta}\lb\frac{2Vr^3\dot{\theta}}{U+J\cos\theta} \pm ir +
\frac{U' \mp iU}{r^2\dot{\theta}}\rb\upr +\e^{\pm 2i\theta}\up{\theta}\n\\
\Gamma_{6\pm} \z \left\{\int\e^{\pm i\theta}\lb 2U'\(1-\frac{U}{r^3\dot{\theta}^2}\)
- 3\(r^2\dot{r}\dot{\theta}
\pm ir^3\dot{\theta}^2-2r^{1/2}V \mp iU\)\rb\d t\right\}\upt\n\\
& & -\e^{\pm i\theta}\lb\frac{2Vr^2\dot{\theta}}{U+J\cos\theta} \(U' \pm
i\(r^3\dot{\theta}^2-U\)\)\(1-
\frac{U}{r^3\dot{\theta}^2}\)\rb\upr\n\\
&& +\e^{\pm i\theta}\lb r^3\dot{\theta}^2-U\rb\up{\theta}
\label{11.12}
\end{eqnarray}
for the original system, \re{11.1}.  In addition there is the symmetry, $\upt $,
 which was used for the reduction of order.

Again we see the very close connection between the structure of
the Ermanno-Bernoulli
 constants and the appropriate variables for the reduction of order.

\section{Conclusions and observations}

For the Kepler problem in three dimensions we obtain the same symmetries as in \re{3.15}
 with the addition of
\begin{equation}
X_{10} = \up{u_3},\label{6.1}
\end{equation}
where $u_3 $ is the azimuthal angle, $\phi $.
Consequently our analysis in the lower dimensional configurational space is
justified by the result that the additional dimension simply adds another
ignorable coordinate to the original system and so a trivial \fode\ to the
reduced system.

In this paper we have examined the process of reduction of order introduced by
Nucci \cite{Nucci} to derive the additional nonlocal symmetries required for
the complete specification of the Kepler problem in the context not only of the
Kepler problem but also in some generalisations which have appeared in the
literature and which possess certain characteristics in common with the Kepler
problem.  In particular we have found that the possession of a conserved vector
similar to that of the Laplace-Runge-Lenz vector, whether or not the magnitude
of the angular momentum is conserved, leads in all cases to a reduced system
consisting of the simple harmonic oscillator and a trivial \fode.  In the
reduced system the Lie point symmetries can be written in a fairly simple
fashion.  When translated to the original system, they are not so simple in
appearance.  However, one can determine the algebraic properties of the several
systems studied from those of the reduced system provided one adds the Lie
point symmetry used in the reduction of order, \viz $\upt $.  The Lie algebra
of Lie point symmetries of the reduced system is $A_1\oplus sl (3,R) $ and
consequently the original systems each have the algebra $2A_1\oplus sl (3,R) $
since the symmetry used in the reduction of order has a zero Lie bracket with
the other symmetries.

Considering the results obtained in this paper we can envisage a reversal of
the procedure.  Instead of taking a system which has a vector of the type of a
Laplace-Runge-Lenz vector we could simply commence with the reduced system and
introduce some transformation of the two ``reduced'' variables and an {\it
Ansatz} on
the relationship defining the variable we have been denoting by $w_3 $.  One
could expect to obtain many ``lame ducks''!  However, there is one aspect which
has the potential for some application.  Many of the systems for which
Laplace-Runge-Lenz vectors have been obtained do not have a known Hamiltonian
representation.  By the procedures of transformations treated in this paper one
could seek to commence with the Hamiltonian of the Kepler problem and find
Hamiltonians for the other systems.  In the case of the system \re {11.2} the
existence of a Hamiltonian has been shown only in the restricted case treated
by Sen.  The attractions for the applications in quantum mechanics are obvious.

In the reduced system the components of the Laplace-Runge-Lenz
vector, the Ermanno-Bernoulli constants, are simply the two
linearly independent first order invariants of the simple harmonic
oscillator.  In fact in the reduced system we have a separation in
the new variables of the Ermanno-Bernoulli constants in the second
order equation and the conservation of a generalised angular
momentum in the first order equation.  We recall that for higher
dimensional oscillators there exist the conserved components of
the Jauch-Hill-Fradkin tensor \cite{Jauch,Fradkin} which play an
important role in the description of the orbit and their time-dependent
counterparts which give the actual trajectory of the particle
\cite{Leach5,Leach6,Lemmer2}. Naturally these
tensors have no role to play in the type of problem considered in
this paper.  However, it is intriguing to ponder the
identity that the corresponding problem of Kepler type would have.  (For a
recent contribution to this more general problem see \cite{Iwai}.)

\section*{Acknowledgements}

MCN thanks the MURST (Cofin 97: Metodi e applicazioni di equazioni
differenziali ordinarie) for its support and PGLL expresses his
deep appreciation of the hospitality of the Dipartimento di
Matematica e Informatica, Universit\`a di Perugia, during the
period in which this work was initiated, and of GEODYSYC,
Department of Mathematics, University of the Aegean, for its
continued hospitality during the period in which this work was
finalised and acknowledges the support of the National Research
Foundation of South Africa and the University of Natal.

\end{document}